\documentclass[conference]{IEEEtran}

\usepackage{cite, makecell}
\usepackage{amsmath,amssymb,amsfonts}
\usepackage{algorithmic}
\usepackage{graphicx}
\usepackage{textcomp}
\usepackage{xcolor, rotating}
\usepackage{balance}
\usepackage{multirow}
\usepackage{gensymb}
\usepackage{array,booktabs}
\usepackage{longtable,tabularx,tabulary}
\usepackage{lipsum}
\usepackage{amssymb,amsmath,epsfig,latexsym, bm}
\usepackage{hyperref}
\hypersetup{colorlinks, pdfa=true, breaklinks, citecolor=blue, urlcolor=black, linkcolor=black}
\usepackage{hyphenat}

\def\BibTeX{{\rm B\kern-.05em{\sc i\kern-.025em b}\kern-.08em
T\kern-.1667em\lower.7ex\hbox{E}\kern-.125emX}}

\newcommand*{\email}[1]{\normalsize\texttt{\href{mailto:#1}{#1}}\par}

\DeclareRobustCommand\onedot{\futurelet\@let@token\@onedot}
\def\@onedot{\ifx\@let@token.\else.\null\fi\xspace}

\makeatother

\begin{document}

\title{Anomaly Detection for Real-World Cyber-Physical Security using Quantum Hybrid Support Vector Machines}

\thanks{978-1-6654-6658-5/22/\$31.00 ©2022 IEEE}


\author{
    \IEEEauthorblockN{%
         Tyler Cultice$^1$,
         Md. Saif Hassan Onim$^1$, Annarita Giani$^2$  and
         Himanshu Thapliyal$^1$ 
                    }

    \IEEEauthorblockA{%
        $^1$Department of Electrical Engineering and Computer Science,\\
        University of Tennessee, Knoxville, TN-37996, United States\\
        $^2$GE Vernova Advanced Research Center, United States}
    
\email{tcultice@vols.utk.edu},
\email{monim@vols.utk.edu},
\email{hthapliyal@utk.edu}
                
    }
\maketitle

\begin{abstract}
Cyber-physical control systems are critical infrastructures designed around highly responsive feedback loops that are measured and manipulated by hundreds of sensors and controllers. Anomalous data, such as from cyber-attacks, greatly risk the safety of the infrastructure and human operators. With recent advances in the quantum computing paradigm, the application of quantum in anomaly detection can greatly improve identification of cyber-attacks in physical sensor data. In this paper, we explore the use of strong pre-processing methods and a quantum-hybrid Support Vector Machine (SVM) that takes advantage of fidelity in parameterized quantum circuits to efficiently and effectively flatten extremely high dimensional data. Our results show an F-1 Score of 0.86 and accuracy of 87\% on the HAI CPS dataset using an 8-qubit, 16-feature quantum kernel, performing equally to existing work and 14\% better than its classical counterpart.
\end{abstract}

\begin{IEEEkeywords}
Anomaly Detection, Support Vector Machine, Cyberphysical System
\end{IEEEkeywords}

\section{Introduction}
Modern industrial and commercial control processes utilize complex, efficiency-driven cyber-physical systems (CPS) that make up critical infrastructures of society. In 2025, it is expected that the CPS market will reach \$9.56 billion and expand by 9.7\% each year \cite{orbis2020}. Despite this growth, the average number of published cyber-physical system vulnerabilities was 115 per month in H2 2022 \cite{claroty2022}. One major target of these attacks is the heart of cyber-physical control systems, the data. CPS utilize data collection to operate physical hardware and improve machine efficiency. Anomalies in this data, especially through cyber-attack, can cause the system to malfunction and cause risk to infrastructural integrity and operator safety. Built-in protections in commonly used systems often give many false positives and false negatives due to their low-dimensional, noisy behavior. As such, CPS systems of the future will continue to increase in complexity, requiring new and innovative ways to identify threats in big data and control systems.

In the cybersecurity paradigm, being able to accurately detect that an attack has occurred in CPS is of utmost importance to protect the integrity of the data and operations involved. Detection of abnormal states through cyberattacks, also known as Anomaly Detection (AD), becomes an increasingly difficult problem as relationships between data features become more hidden from view. However, quantum allows us to efficiently compute high-dimensional correlation between features, providing SVMs with dimensionally-compressed data for easier classification. While many anomaly detection systems have been proposed in the past, few works focus on exploring Quantum-assisted SVM application in anomalies of real-world cyber-physical datasets.

The increase in public accessibility of quantum computing through Noisy Intermediate-Scale Quantum (NISQ) computers has encouraged researchers to look into quantum-assisted machine learning tasks. Ideally, quantum algorithms allow for us to find more complex patterns or correlations in data that would otherwise be extremely computationally expensive for classical processors. In particular, quantum fidelity kernels are of interest in domains like anomaly detection for creating a highly correlated kernel matrix between data vectors that can be directly fed into Support Vector Machines (SVM). Using these quantum properties in machine learning applications such as this problem can greatly improve the ability to represent patterns in high dimensional data more accurately in kernel-method machine learning algorithms.

Effective use of quantum machine learning (QML) is a highly sought research topic as of today, with no conclusion in sight as to how it will perform compared to future classical models. Additionally, with the current limitation and bottleneck of quantum resources, constraints to training data must be considered that can greatly reduce triviality of quantum machine learning models. A lot of emphasis in real QML application is around pre-processing and post-processing of the data to maximize performance and minimize the overhead required for high-feature datasets. Finding the best methods to encode, process, and utilize classical data while adhering the modern quantum computing paradigm's limitations is key to real-world application of QML-based anomaly detection.

Our contributions in this article can be summarized as the following:
\begin{itemize}
    \item We utilize a hybrid quantum-classical SVM approach to Anomaly Detection to investigate quantum SVMs in real-world CPS data.
    \item We explore unique, effective data pre-processing methods to improve the performance of QML for classical data.
    \item We demonstrate an accuracy of up to 87\% on the HAI 20.07 dataset using a quantum kernel, providing a 14\% increase in accuracy from similar classical SVMs. 
\end{itemize}

Our organization is as follows: Section~\ref{Background} is a background on quantum encoding, SVMs, and related AD work. Section~\ref{Dataset} describes the HAI dataset and features. Section~\ref{Methodology} details the design of the hybrid quantum-kernel SVM. Section~\ref{Result} provides the SVM results with the HAI dataset, including comparison to classical SVMs and related works. Section~\ref{Discussion} provides a conclusion and discussion of future work.

\section{Background}
\label{Background}
\subsection{Embedding Classical Data to Quantum}
In QML, data encoding is performed near the beginning the circuit to initialize each qubit with a feature of the classical data point, resulting in a quantum state encoded with $N$ features. This is generally done through angle encoding, which rotates each qubit by the values of the features. In dense angle encoding, two features can be mapped to each qubit by taking advantage of the relative phase degree of freedom \cite{LaRose2020}, resulting in $N$ features expressed by $n=N/2$ qubits. The feature vector $x = [x_1,...,x_N]^T \in \mathbb{R}^N$ is mapped using $S_x$, a set of rotation gates at the start of $U(x)$. This will prepare our state for the later unitary operations of the feature map.


\subsection{Quantum Fidelity Kernels}
Typically, Quantum Support Vector Machines (QSVM) work under a hybrid quantum-classical design approach, using a classical SVM paired with a quantum kernel. Similar to classical SVMs, a kernel function is utilized that measures the state overlap, or fidelity, between two quantum states produced by a parameterized "feature map". This state fidelity is defined in equation \ref{eq:fidelityDef}. The feature map encodes data-points to quantum states within an expressive $n$-qubit Hilbert space $\mathcal{H}$, flattening the dimensionality of the data into an SVM-separable state \cite{Schuld2019}. In most feature maps, data is angle (or dense-angle) encoded into initialized base-state qubits, or $|0\rangle^{\otimes n}$. Various transformations, such as controlled gates (C-NOT) for entanglement, are performed which we define as $G(x)$. This results in a output state, $\phi(x)$, used for fidelity.

\begin{equation}
K(x,y) = |\langle\phi(x)|\phi(y)\rangle|^2
\label{eq:fidelityDef}
\end{equation}

\begin{figure}[]
\centering
\includegraphics[width=\columnwidth]{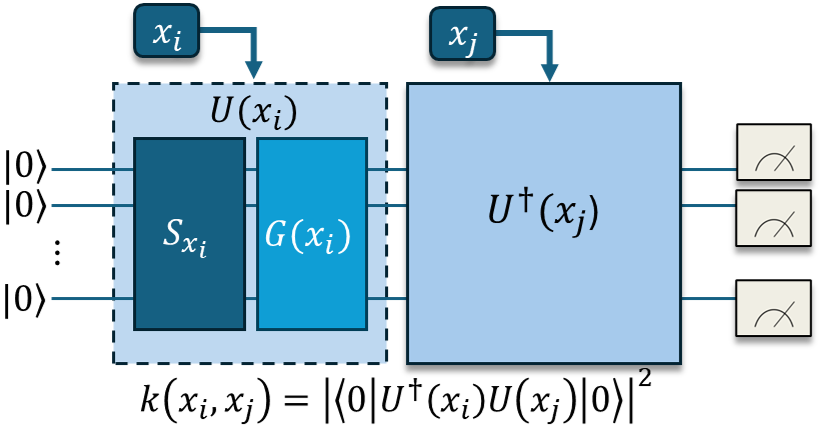}
\caption{Hilbert-Schmidt inner product calculation for determining the fidelity of two feature-mapped values in the kernel, ($x_i$,$x_j$). A unitary $U(x_i)$ is performed, followed by its inverse $U^\dag(x_j)$. $S_{x_i}$ refers to the encoding circuit, while $G(x_i)$ refers to the unitary feature map operations. This calculation is repeated for all elements of $k(x_i,x_j)$ to establish the inner product matrix.}
\label{fig:computeUncompute}
\end{figure}

A qubit-efficient way of finding the fidelity without creating two separate qubit states is the Hilbert-Schmidt Inner-Product method, as described by Havlíček et al. \cite{Havlicek2019}. This process is shown in Figure \ref{fig:computeUncompute}. In this design, a unitary $U(x_i)$ is applied, followed by its inverse $U^\dag(x_j)$. The measured frequency of the result being the base state (usually $|0\rangle$) defines how similar $x_i$ and $x_j$ are. For example, the resulting state from $U(x_i)$ would be perfectly inverted by $U^\dag(x_j)$ in an ideal situation if $x_i = x_j$. When repeated for all $k(x_i,x_j)$, it results in an inner-product matrix of quasi-probabilities bounded within $[0,1]$.

\subsection{One-Class Support Vector Machines}
Support Vector Machines (SVM) are a supervised learning technique originally introduced by Boser et al.~\cite{svm}. Used for classification and regression problems, they determine the best hyperplane to split data points into distinct classes in a high-dimensional space. The hyperplane is set to maximize the margin between the nearest data points of distinct classes, hence improving the model's generalization and robustness. One-Class SVM (OCSVM) is a variation of SVM that is intended for anomaly or novelty detection assignments. Unlike regular SVMs, which are trained on data with labeled classes, One-Class SVMs are trained on datasets that only contain one class, usually the normal (non-anomaly) or majority class. The OCSVM learns to determine the distribution of this class and then detects deviations or outliers during prediction.

\subsection{Related Work in Anomaly Detection}
Anomaly detection is a highly explored topic in classical machine learning. Work by Tushkanova et al. \cite{Tushkanova2023} highlighted classical SVMs performance compared to other existing AD methods applied to various CPS datasets, including SWaT and HAI. Additionally, a comparative study was conducted on ICS data by Kim et al. \cite{Kim2023} that deeply described the relationship of features in HAI and analyzed the distributions of data.

Quantum machine learning has given rise to popularity in Quantum Anomaly Detection (QAD) due to the benefits of quantum. Quantum autoencoders \cite{Ngairangbam2022}, "Circuit Learning" \cite{Alvi2023}, and kernel SVMs \cite{woniak2023, Schuhmacher2023} are proposed methods of detecting anomalies in Large Hadron Colliders (LHC) and other new physics phenomena. Some work has explored the use of classical and quantum kernel SVMs in synthetic datasets \cite{tscharke2023} and SCADA wind turbine systems \cite{jullian2022}. However, little has been done in detecting anomalies in cyber-physical systems. 

\section{Dataset: HIL-based Augmented ICS (HAI)}
\label{Dataset}
\begin{figure}[]
\centering
\includegraphics[width=\columnwidth]{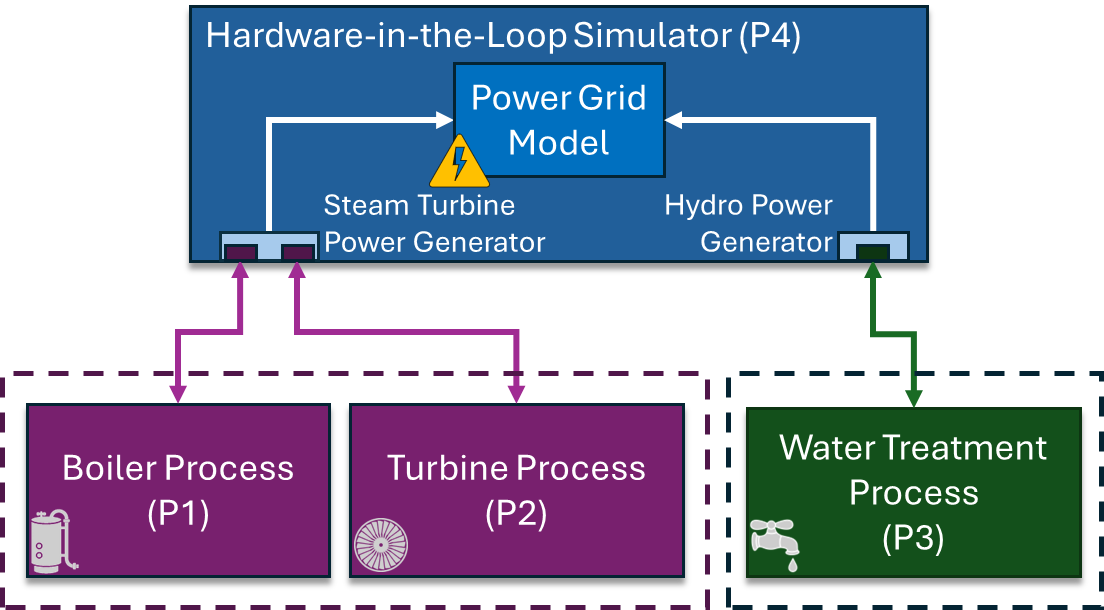}
\caption{HIL-based Augmented ICS (HAI) structure for steam-turbine and hydro-power. P1 (Boiler), P2 (Turbine), and P3 (Water Treatment) are all connected by P4, the overall controller that simulates a power grid model.}
\label{fig:haiStructure}
\end{figure}

An essential challenge to identifying the suitability of quantum machine learning on data of thousands of cyber-physical sensors is to select suitable real-world datasets. In this work, we selected the Hardware-in-Loop-based (HIL) Augmented Industrial Control System (ICS) Security Dataset, or HAI 20.07 \cite{hai2020}. The HAI dataset is a realistic ICS testbed that physically emulates hydro-power and steam-turbine power generation using Programmable Logic Controllers (PLCs). The system contains four critical processes that are extremely essential to safe operation: the Boiler process (P1), the Turbine process (P2), the Water Treatment process (P3), and the Hardware-in-Loop Simulation (P4). How these processes are interconnected can be seen in Figure \ref{fig:haiStructure}. Input parameters are set for feedback control loops which alter the behavior of the system and PID controllers, including water pressure, flow-rate, boiling temperature, and expected power output.

In this dataset, there are 59 features (including the input parameters) spread across all four processes. These features are generally measured points within each process that directly impact how the system will respond, react, and possibly warn the operator. Training data was captured from normal operation with no anomalies for 177 hours, with various set-points throughout the 24-hour period. The test data contains both normal and anomalous data captured over 123 hours with similar inputs. In total, 38 attacks were conducted, including 14 different attack primitives and combinations of attacks. Anomalies are defined as attacker modified data showing significant deviation from typical patterns between features. The resolution of the dataset is at 1 second each. The percentage of anomalies in the test data we used is 4\% \cite{Tushkanova2023}, requiring additional considerations for the class imbalance.

\section{Methodology}
\label{Methodology}
\begin{figure*}[]
\centering
\includegraphics[width=0.9\textwidth]{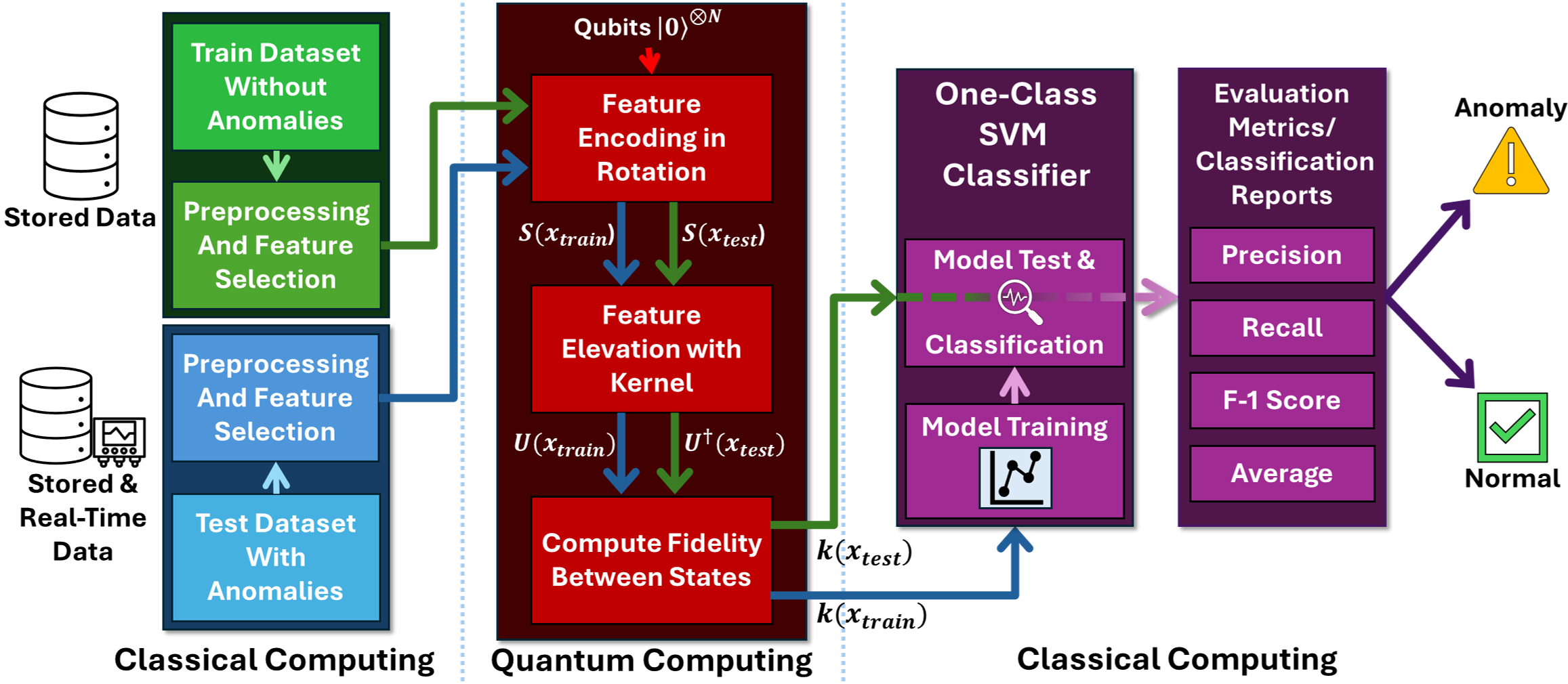}
\caption{Pipeline of the hybrid-quantum SVM anomaly detector. Pre-processing occurs classically, is turned into a fidelity inner-product kernel using a quantum computer, then utilized for SVM anomaly detection. Data kernels can then be passed into the trained SVM for detection of anomalies.}
\label{fig:overallFlow}
\end{figure*}

The general flow of this hybrid SVM approach is as follows: perform feature selection and pre-processing, calculate the quantum Hilbert-Schmidt inner product between the current data vector and the training data, then use it to train and classify on the SVM's hyperplane. This flowchart can be seen in Figure \ref{fig:overallFlow}. The SVM-based design of this approach takes advantage of quantum fidelity to map the feature space and find similarities in data. The pre-processing steps and SVM are performed classically before and after the parameterized quantum inner-matrix.

\subsection{Data Preprocessing and Feature Selection}
This stage focuses on preparing the data for the feature selection and SVM techniques. Firstly, the moving average of 60 overlapping sample windows helps to reduce noise. It also captures the time series trends following the Equation~\eqref{sma}. Here $w$ is the window length and $x_i$ is the sample at time $i$.

\begin{equation}
    ma_t = \frac{1}{w} \times \sum_{i=t-w+1}^t x_i
    \label{sma}
\end{equation}

Next, the categorical samples are replaced with their numerical histogram values which are more fitted for the ML algorithms. Finally, the dataset is standardized with standard-scaler transformation or Z-score normalization with Equation~\eqref{std}. Here, $\mu$ is the sample mean and $\sigma$ is the standard deviation. This moves the mean of the distribution to zero and the standard deviation to 1. This helps to equalize the scales of different units. Also, standardization tends to bring faster convergence for optimization techniques like gradient descent.
\begin{equation}
    x_{std} = \frac{x_o - \mu}{\sigma}
    \label{std}
\end{equation}

The pre-processing steps end with the feature selection from the feature importance rank calculated from a decision tree by the Gini impurity. Based on the relative decrease in Gini impurity the features are ranked from most important to less. The number of selected feature is chosen based on the number of qubits and encoded features per qubit. 


\subsection{Fidelity Kernel}
In order to find high-dimensional similarities between features of the data, we utilize a kernel based on the quantum fidelity between feature-encoded entangled states of each data point. In this kernel, a parameterized feature map is utilized to find similarities between the classical features in a higher-dimensional quantum space. 

\begin{figure}[]
\centering
\includegraphics[width=\columnwidth]{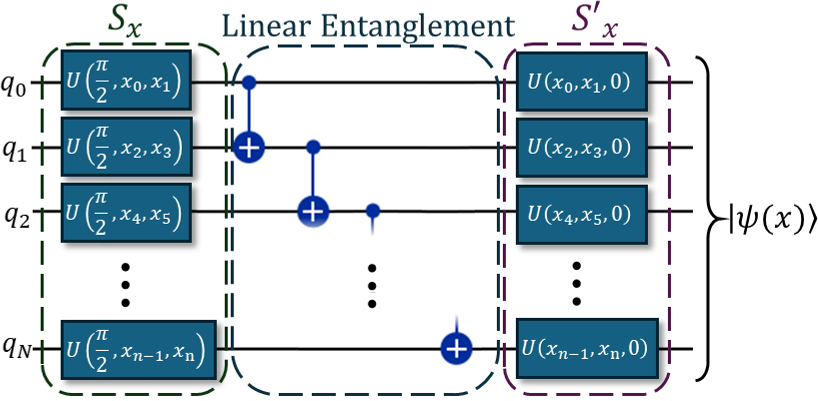}
\caption{The feature map. $N=n/2$ qubits are used to encode $n$ features. Data is encoded using generic rotation gates, or U gates, then entangled and rotated further. This circuit may be repeated multiple times (i.e., 3 times) to drive further interaction between qubits and features.}
\label{fig:featureMap}
\end{figure}

The feature map can be seen in Figure \ref{fig:featureMap}. Inspired by Woźniak et al. \cite{woniak2023}, it contains an encoding ansatz using two degrees of rotational freedom with nearest neighbor-entanglement and a repeated rotation. We repeat the feature map three times in $U(x)$ to drive more interaction between features and correlations. The data is normalized to $[-\pi,\pi]$ to properly map the data to the bounds of the angle embedding operation. The quantum fidelity operation is repeated for every combination of data samples, or $O(\lvert x_{input} \rvert \times \lvert x_{train} \rvert \times shots)$, resulting in a matrix kernel.

During training, the calculated kernel is a self inner product of the training data. To optimize the self inner product, we assume the kernel to be symmetric, or $k(x,y) = k(y,x)$, allowing us to cut the number of circuit runs by more than half. For the testing set (and real-time operation), an inner product kernel between the testing data and training data is calculated. Lower fidelity similarity is expected from the anomalous testing samples and higher for normal samples. The resulting kernels are post-processed by exponentiating the values, $k_{p}(x_i,x_j)=e^{k(x_i,x_j)}$. This exponentiation further magnifies the differences between kernel elements.

\subsection{Training the One-Class Support Vector Classifier}
The goal of the training phase is to identify the ideal hyperplane with the widest margin that divides the data points of various classes. In kernel-based SVMs, the optimization issue is solved in the feature space generated by the kernel function of choice. Usually, the optimization problem is expressed as a quadratic programming (QP) problem and resolved by applying a convex optimization solver. Additionally, the data points that are closest to the hyperplane of the decision boundary are known as the support vectors, which are also learned by SVM during training and define the decision boundary. Given a set of training data from only one class $\{x_i\}_{i=1}^N$, the decision function for a One-Class SVM can be represented as Equation~\eqref{obj_func}.

\begin{equation}
    f(x) = sign\left( \sum_{i=1}^N\sum_{j=1}^N  \alpha_{ij} \times k(x_i, x_j) + b - \rho \right)
    \label{obj_func}
\end{equation}

Here, $k(x_i, x_j)$ is the kernel function that computes the similarity between the input vectors $x_i$ and $x_j$ in the feature space. $\alpha_{ij}$ are the multipliers associated with the training samples. $b$ is the bias term and $\rho$ is the threshold. Now, the optimization is done to maximize the margin between the decision boundary represented as Equation~\eqref{opt} for a given constraint mentioned in Equation~\eqref{given_cons}.

\begin{equation}
    \underset{\alpha, \rho}{maximize} = \frac{1}{2} \times ||\omega||^2 - \rho + \frac{1}{\nu \times N} \sum_{i=1}^N{\zeta_i}
    \label{opt}
\end{equation}

\begin{equation}
    when, \omega^T \times \phi(x_i) + b \geq \rho - \zeta_i
    \label{given_cons}
\end{equation}

Here, $\omega$ is the vector normal to the decision hyperplane, $\phi(x_i)$ is the basis function, $\zeta$ is the relaxation value, and $\nu$ controls the trade-off between maximizing the margin and controlling the fraction of outliers. This optimization problem is solved iteratively with gradient descent or sequential minimal optimization. In our work, we used \textit{scikit-learn} module's \textit{svm} class to implement this.

\subsection{Testing and Anomaly Detection}
After training, the AD model will have identified decision boundaries to classify anomalous and normal data points. When utilizing the model for anomaly detection, the input data, such as testing data, creates an inner product matrix with the training data, $x_{test} \times x_{train}$. This matrix is used to map the points and classify them on the hyperplane.

\begin{figure}[]
\centering
\includegraphics[width=.98\columnwidth]{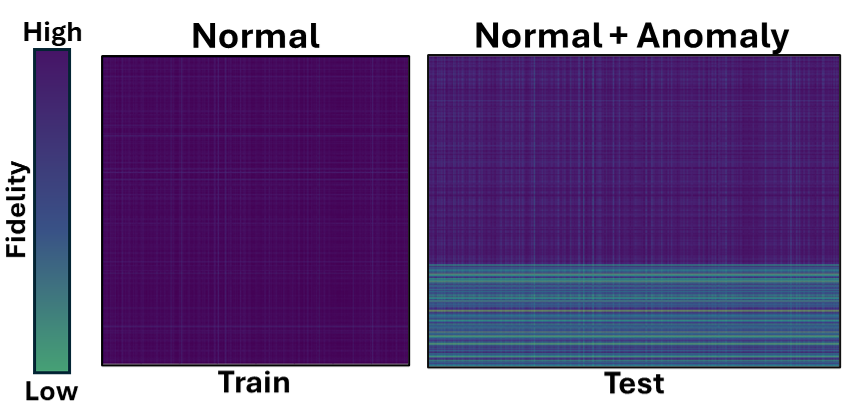}
\caption{Example of the 8-qubit fidelity kernel design on the HAI dataset for the train and test sets. Lower fidelity (green) means higher likelihood of the data being decided as anomalous. This can be seen in the test set kernel, where 1000 normal samples are provided followed by 500 anomalous samples. }
\label{fig:kernelExample}
\end{figure}

\section{Result Analysis}
\label{Result}
The best way to identify the effectiveness of quantum kernels in CPS is to utilize realistic data captured on real hardware. While it brings forth additional challenges compared to synthetic data, we believe this approach gives a much better analysis of QAD in the real world. Thus, we utilized the HAI 20.07 dataset to validate our design. To identify the ideal quantum benefit, noise is not factored into this work. Quantum computing operations in this design were simulated using IBM Qiskit \cite{qiskit} with no noise model. This section describes the results we achieved with the HAI dataset and compares it to equivalent classical models and existing AD literature.

\subsection{Impact of Quantum Mapping vs. Classical}
The value of the kernel elements are determined by the overlap, or fidelity, of quantum states initialized by the features. As seen in Figure \ref{fig:kernelExample}, the more distance the $N$ features are between $x_i$ and $x_j$ of $k(x_i,x_j)$, the lower the fidelity, $x\rightarrow 0$. In the training inner matrix kernel ($x_{train} \times x_{train}$), the data mostly has a strong fidelity to one another ($x\rightarrow1$), with variation in certain data point relationships. However, when introducing anomalies to the end of the test set ($x_{test} \times x_{train}$), the fidelity of these samples significantly decreases compared to the normal samples.

To properly identify the benefits of using a fidelity-based quantum kernel, a table of various performance metrics is provided in Table \ref{table:performance}. The number of qubits is tied to the number of features, giving both pros and cons to increasing or decreasing the number of qubits and features. While more qubits would generally increase expressivity, this particular dataset favored a feature count of 16. This is situational to both the dataset structure and the behavior of the quantum computation. Thus, we can expect a situational balance between the dataset features and size of quantum circuit that the user must explore to find the optimal parameters for good AD performance.

\begin{table}
\caption{Performance metrics of Quantum SVM model and typical classical SVM based on feature (and qubit) counts. }
\label{table:performance}
\centering
\resizebox{\columnwidth}{!}{
    \setlength{\tabcolsep}{3pt}
    \renewcommand*{\arraystretch}{1.15}
    \begin{tabular}{cccccc}
    \toprule
     \bf Method & \bf Features & \bf Accuracy & \bf Precision & \bf Recall & \bf F1 \\
        \cmidrule{1-6}
        & 8 (4-Qubits) & 86\% & 0.87 & 0.86 & 0.85 \\
        \cmidrule{2-6}
        \bf Quantum & \bf 16 (8-Qubits) & \bf 87\% & \bf 0.88 & \bf 0.87 & \bf 0.86 \\
        \cmidrule{2-6}
        & 24 (12-Qubits) & 82\% & 0.82 & 0.82 & 0.82 \\
        \midrule
        & 8 & 74\% & 0.81 & 0.74 & 0.67 \\
        \cmidrule{2-6}
        \bf Classical & 16 & 73\% & 0.81 & 0.73 & 0.65 \\
        \cmidrule{2-6}
        & 24 & 72\% & 0.79 & 0.72 & 0.65 \\
        \bottomrule
    \end{tabular}
    }
\end{table}

Overall, our design works best on the HAI dataset with 16 features and 8 qubits with 87\% accuracy and F-1 score of 0.86. Comparing this to a classical counterpart with same features, we can see a 14\% average increase in accuracy and 0.21 increase in F-1 score between the models. The classical kernel struggles to identify anomalies (Recall) and succeeds at avoiding false positives (Precision), whereas the quantum kernel is great at identifying both (high F1 and accuracy).

\subsection{Comparison to Existing CPS Solutions}
To identify how our performance matches up against other quantum AD methods and datasets, we present in this section a comparison of ML metrics to existing literature in QAD. We selected the closest parameterized results from existing literature to our own analysis for a more accurate and fair comparison. Additionally, with fewer works done with the HAI dataset, we have included results of existing work using similar CPS datasets in Table \ref{table:comparison}.

\begin{table}
\caption{Comparison of our work with existing quantum literature.}
\centering
\label{table:comparison}
\resizebox{1.01\columnwidth}{!}{
    \setlength{\tabcolsep}{1.5pt}
    \renewcommand*{\arraystretch}{1.15}
    \begin{tabular}{cccccc}
    \toprule
     \makecell[c]{\bf Author and Year} & \makecell[c]{\bf QML\\ \bf Algorithm} & \bf Dataset & \makecell[c]{\bf \# of \\ \bf Qubits} & \bf Dimensions & \bf Metrics\\
        \midrule
        \makecell[c]{Gouveia~et al.~\cite{network2}\\ 2020} & QSVM & \makecell[c]{NF-UNSW-\\NB15} & 16 & 16 & \makecell[c]{Acc: 64}\\
        \midrule
        \makecell[c]{Correa-Jullia~et al.~\cite{jullian2022}\\ 2022} & QSVM & WTS & 8 & 16 & \makecell[c]{Acc: 88.8\% \\ F1: 0.893}\\
        \midrule
        \makecell[c]{Tscharke~et al.~\cite{tscharke2023}\\ 2023} & QSVR & KDD & 5 & 5 & \makecell[c]{Acc: 82.0\% \\ F1: 0.78}\\
        \midrule
        \makecell[c]{Wang~et al.~\cite{MNIST}\\ 2023} & QHDNN & \makecell[c]{Fashion\\MNIST} & 16 & 16 & \makecell[c]{AUC: 88.24}\\
        \midrule
        \makecell[c]{Wang~et al.~\cite{MNIST}\\ 2023} & QHDNN & MNIST & 16 & 16 & \makecell[c]{AUC: 89.41}\\
        \midrule
        \makecell[c]{Kukliansky~et al.~\cite{network}\\ 2024} & QNN & \makecell[c]{NF-UNSW-\\NB15} & 16 & 16 & \makecell[c]{F1: 0.86}\\
        \midrule

        \makecell[c]{\bf Our Work\\ \bf 2024} & QOC-SVM & \bf HAI & \bf 8 & \bf 16 & \makecell[c]{\bf Acc: 87.0\% \\ \bf F1: 0.86}\\
        \bottomrule
    \end{tabular}
    }
\end{table}

It can be seen that our work performs similarly to existing highly-tuned, time-series solutions for CPS datasets. This is impressive due to the time-series nature of the HAI dataset, which favors solutions like neural networks. We also compared this to QML anomaly detection in spatial dimensions tested on image datasets that required Quantum Hybrid Deep Neural Networks (QHDNN) or Quantum Neural Networks (QNN) for feature extraction. The trade-off for the NN's high number of trainable parameters resulted in almost similar metrics as our preprocessing Quantum kernel with fewer parameters.


\section{Discussion and Future Work}
\label{Discussion}
In this work, we introduced a method of anomaly detection in CPS systems based around the quantum advantage for high-dimensional computation of data. By utilizing a kernel prepared from the fidelity of parameterized quantum circuits, this work flattens the higher dimensional feature space of real-world data for effective SVM classification. Our selection of pre-processing methods and a robust feature map provided strong improvement over existing methods. For the real-world power CPS data, HAI 20.07 \cite{hai2020}, we see an F-1 score of 0.86 and 87\% accuracy with 16 features (8 qubits), performing 14\% better than similar classical solutions and equally to quantum solutions. This is an improvement beyond statistical variation/error for our quantum solution.

While these results demonstrate that detection of anomalies can be achieved on real world CPS data, quantum kernel computation can be an expensive task due to the nature of NISQ computers. Quantum kernels utilize $O(\lvert x_{input} \rvert \times \lvert x_{train} \rvert \times shots)$ individual parameterized $N$-qubit circuits just to create the training or testing kernel. Due to this, real-time systems like CPS can't get second or minute AD resolution because of current architecture, noise, and qubit limitations. Future work should explore minimizing the overhead using: circuit parallelization \cite{Ohkura2022}, batching of real-time data to maximize QPU usage \cite{Domeniconi2001}, and quantum-classical ensemble methods \cite{Araya2017}. Both classically and quantumly improving the quantum pipeline should be explored in further research, including expansion of this work to other CPS datasets.

Furthermore, our no-noise simulation resulted in highly distinguishable fidelity differences between the two classes. However, in reality, NISQ computers are notorious for noisy and error-prone calculations. Machine learning heavily relies on high-precision and accuracy in values, causing many performance issues in quantum kernels when exposed to too much real-hardware error. Further exploration of the effects of noise, noise mitigation methodologies, etc. remains of utmost importance to improve NISQ anomaly detection performance.

\section{Acknowledgment}
This material is based upon work supported by the National Science Foundation Graduate Research Fellowship under Grant No. 1938092.

\bibliographystyle{style}
\bibliography{reference}

\begin{thebibliography}{10}
\providecommand{\url}[1]{#1}
\csname url@samestyle\endcsname
\providecommand{\newblock}{\relax}
\providecommand{\bibinfo}[2]{#2}
\providecommand{\BIBentrySTDinterwordspacing}{\spaceskip=0pt\relax}
\providecommand{\BIBentryALTinterwordstretchfactor}{4}
\providecommand{\BIBentryALTinterwordspacing}{\spaceskip=\fontdimen2\font plus
\BIBentryALTinterwordstretchfactor\fontdimen3\font minus \fontdimen4\font\relax}
\providecommand{\BIBforeignlanguage}[2]{{%
\expandafter\ifx\csname l@#1\endcsname\relax
\typeout{** WARNING: IEEEtran.bst: No hyphenation pattern has been}%
\typeout{** loaded for the language `#1'. Using the pattern for}%
\typeout{** the default language instead.}%
\else
\language=\csname l@#1\endcsname
\fi
#2}}
\providecommand{\BIBdecl}{\relax}
\BIBdecl

\bibitem{orbis2020}
O.~Research, ``Global cyber physical system market 2020 by company, regions, type and application, forecast to 2025,'' https://www.orbisresearch.com/reports/index/2015-2025-global-cyber-physical-system-cps-market-research-by-type-end-use-and-region, 2020.

\bibitem{claroty2022}
B.~Ofner, R.~Mesika, N.~Erez, S.~Brizinov, and C.~Fradkin, ``{State of XIoT Security: 2H 2022},'' Claroty, Tech. Rep., H2 2022.

\bibitem{LaRose2020}
\BIBentryALTinterwordspacing
R.~LaRose and B.~Coyle, ``Robust data encodings for quantum classifiers,'' \emph{Phys. Rev. A}, vol. 102, p. 032420, Sep 2020. [Online]. Available: \url{https://link.aps.org/doi/10.1103/PhysRevA.102.032420}
\BIBentrySTDinterwordspacing

\bibitem{Schuld2019}
\BIBentryALTinterwordspacing
M.~Schuld and N.~Killoran, ``Quantum machine learning in feature hilbert spaces,'' \emph{Phys. Rev. Lett.}, vol. 122, p. 040504, Feb 2019. [Online]. Available: \url{https://link.aps.org/doi/10.1103/PhysRevLett.122.040504}
\BIBentrySTDinterwordspacing

\bibitem{Havlicek2019}
\BIBentryALTinterwordspacing
V.~Havl{\'i}{\v{c}}ek, A.~D. C{\'o}rcoles, K.~Temme, A.~W. Harrow, A.~Kandala, J.~M. Chow, and J.~M. Gambetta, ``Supervised learning with quantum-enhanced feature spaces,'' \emph{Nature}, vol. 567, no. 7747, pp. 209--212, Mar 2019. [Online]. Available: \url{https://doi.org/10.1038/s41586-019-0980-2}
\BIBentrySTDinterwordspacing

\bibitem{svm}
B.~E. Boser, I.~M. Guyon, and V.~N. Vapnik, ``A training algorithm for optimal margin classifiers,'' in \emph{Proceedings of the fifth annual workshop on Computational learning theory}, 1992, pp. 144--152.

\bibitem{Tushkanova2023}
\BIBentryALTinterwordspacing
O.~Tushkanova, D.~Levshun, A.~Branitskiy, E.~Fedorchenko, E.~Novikova, and I.~Kotenko, ``Detection of cyberattacks and anomalies in cyber-physical systems: Approaches, data sources, evaluation,'' \emph{Algorithms}, vol.~16, no.~2, 2023. [Online]. Available: \url{https://www.mdpi.com/1999-4893/16/2/85}
\BIBentrySTDinterwordspacing

\bibitem{Kim2023}
\BIBentryALTinterwordspacing
B.~Kim, M.~A. Alawami, E.~Kim, S.~Oh, J.~Park, and H.~Kim, ``A comparative study of time series anomaly detection models for industrial control systems,'' \emph{Sensors}, vol.~23, no.~3, 2023. [Online]. Available: \url{https://www.mdpi.com/1424-8220/23/3/1310}
\BIBentrySTDinterwordspacing

\bibitem{Ngairangbam2022}
\BIBentryALTinterwordspacing
V.~S. Ngairangbam, M.~Spannowsky, and M.~Takeuchi, ``Anomaly detection in high-energy physics using a quantum autoencoder,'' \emph{Phys. Rev. D}, vol. 105, p. 095004, May 2022. [Online]. Available: \url{https://link.aps.org/doi/10.1103/PhysRevD.105.095004}
\BIBentrySTDinterwordspacing

\bibitem{Alvi2023}
\BIBentryALTinterwordspacing
S.~Alvi, C.~W. Bauer, and B.~Nachman, ``Quantum anomaly detection for collider physics,'' \emph{Journal of High Energy Physics}, vol. 2023, no.~2, p. 220, Feb 2023. [Online]. Available: \url{https://doi.org/10.1007/JHEP02(2023)220}
\BIBentrySTDinterwordspacing

\bibitem{woniak2023}
K.~A. Woźniak, V.~Belis, E.~Puljak, P.~Barkoutsos, G.~Dissertori, M.~Grossi, M.~Pierini, F.~Reiter, I.~Tavernelli, and S.~Vallecorsa, ``Quantum anomaly detection in the latent space of proton collision events at the lhc,'' \emph{arXiv preprint arXiv:2301.10780}, 2023.

\bibitem{Schuhmacher2023}
\BIBentryALTinterwordspacing
J.~Schuhmacher, L.~Boggia, V.~Belis, E.~Puljak, M.~Grossi, M.~Pierini, S.~Vallecorsa, F.~Tacchino, P.~Barkoutsos, and I.~Tavernelli, ``Unravelling physics beyond the standard model with classical and quantum anomaly detection,'' \emph{Machine Learning: Science and Technology}, vol.~4, no.~4, p. 045031, nov 2023. [Online]. Available: \url{https://dx.doi.org/10.1088/2632-2153/ad07f7}
\BIBentrySTDinterwordspacing

\bibitem{tscharke2023}
K.~Tscharke, S.~Issel, and P.~Debus, ``Semisupervised anomaly detection using support vector regression with quantum kernel,'' in \emph{2023 IEEE International Conference on Quantum Computing and Engineering (QCE)}, vol.~01, 2023, pp. 611--620.

\bibitem{jullian2022}
\BIBentryALTinterwordspacing
C.~Correa-Jullian, S.~Cofre-Martel, G.~San~Martin, E.~Lopez~Droguett, G.~de~Novaes Pires~Leite, and A.~Costa, ``Exploring quantum machine learning and feature reduction techniques for wind turbine pitch fault detection,'' \emph{Energies}, vol.~15, no.~8, 2022. [Online]. Available: \url{https://www.mdpi.com/1996-1073/15/8/2792}
\BIBentrySTDinterwordspacing

\bibitem{hai2020}
H.-K. Shin, W.~Lee, J.-H. Yun, and H.~Kim, \emph{HAI 1.0: HIL-Based Augmented ICS Security Dataset}.\hskip 1em plus 0.5em minus 0.4em\relax USA: USENIX Association, 2020.

\bibitem{qiskit}
{Qiskit contributors}, ``Qiskit: An open-source framework for quantum computing,'' 2023.

\bibitem{network2}
A.~Gouveia and M.~Correia, ``Towards quantum-enhanced machine learning for network intrusion detection,'' in \emph{2020 IEEE 19th International Symposium on Network Computing and Applications (NCA)}.\hskip 1em plus 0.5em minus 0.4em\relax IEEE, 2020, pp. 1--8.

\bibitem{MNIST}
\BIBentryALTinterwordspacing
M.~Wang, A.~Huang, Y.~Liu, X.~Yi, J.~Wu, and S.~Wang, ``A quantum-classical hybrid solution for deep anomaly detection,'' \emph{Entropy}, vol.~25, no.~3, 2023. [Online]. Available: \url{https://www.mdpi.com/1099-4300/25/3/427}
\BIBentrySTDinterwordspacing

\bibitem{network}
A.~Kukliansky, M.~Orescanin, C.~Bollmann, and T.~Huffmire, ``Network anomaly detection using quantum neural networks on noisy quantum computers,'' \emph{IEEE Transactions on Quantum Engineering}, vol.~5, pp. 1--11, 2024.

\bibitem{Ohkura2022}
Y.~Ohkura, T.~Satoh, and R.~Van~Meter, ``Simultaneous execution of quantum circuits on current and near-future nisq systems,'' \emph{IEEE Transactions on Quantum Engineering}, vol.~3, pp. 1--10, 2022.

\bibitem{Domeniconi2001}
C.~Domeniconi and D.~Gunopulos, ``Incremental support vector machine construction,'' in \emph{Proceedings 2001 IEEE International Conference on Data Mining}, 2001, pp. 589--592.

\bibitem{Araya2017}
\BIBentryALTinterwordspacing
D.~B. Araya, K.~Grolinger, H.~F. ElYamany, M.~A. Capretz, and G.~Bitsuamlak, ``An ensemble learning framework for anomaly detection in building energy consumption,'' \emph{Energy and Buildings}, vol. 144, pp. 191--206, 2017. [Online]. Available: \url{https://www.sciencedirect.com/science/article/pii/S0378778817306904}
\BIBentrySTDinterwordspacing

\end{thebibliography}

\end{document}